\documentclass[12pt]{iopart}

\begin{document}

\maketitle

\title{Special Functions of Mathematical Physics: A Unified Lagrangian Formalism}

\author{Z. E. Musielak,  N. Davachi and M.  Rosario-Franco}
\address{Department of Physics, The University of Texas at 
Arlington, Arlington, TX 76019, USA}

\begin{abstract}
Lagrangian formalism is established for differential equations with
special functions of mathematical physics as solutions.  Formalism is based 
on either standard or non-standard Lagrangians.  This work shows that the 
procedure of deriving the standard Lagrangians leads to Lagrangians for 
which the Euler--Lagrange equation vanishes identically, and that only some of
these Lagrangians become the null Lagrangians with the well-defined gauge 
functions.  It is also demonstrated that the non-standard Lagrangians require 
that the Euler--Lagrange equations are amended by the auxiliary conditions, which
is a new phenomenon in the calculus of variations.  The~existence of the auxiliary 
conditions has profound implications on the validity of the Helmholtz conditions.  
The obtained results are used to derive the Lagrangians for the Airy, Bessel, 
Legendre and Hermite equations.  The presented examples clearly demonstrate 
that the developed Lagrangian formalism is applicable to all considered differential 
equations, including the Airy (and other similar) equations, and that the regular 
and modified Bessel equations are the only ones with the gauge functions.  
Possible implications of the existence of the gauge functions for these 
equations are~discussed.
\end{abstract}


\section{Introduction}

There are numerous applications of linear, second-order ordinary differential
equations (ODEs) in applied mathematics and physics [1,2].  The most commonly 
used are the ODEs whose solutions are given by the special functions (SFs) of 
mathematical physics defined in [3,4,5].   In this paper, we concentrate on these 
ODEs and introduce $\mathcal{Q}_{sf}$ to be a set of such equations.
  
Let $\hat D =  {{d^2} / {dx^2}} + B(x) {{d} / {dx}} + C(x)$ be a linear 
operator with $B(x)$ and $C(x)$ being ordinary (with the maps $B :\ \mathcal{R} 
\rightarrow \mathcal{R}$ and $C :\ \mathcal{R} \rightarrow \mathcal{R}$, with 
$\mathcal{R}$ denoting the real numbers) and smooth with at least two continuous 
derivatives ($\mathcal{C}^{2}$) functions defined either over a restricted interval
$(a, b)$ or an infinite interval $(- \infty , \infty)$ depending on the ODE of 
$\mathcal{Q}_{sf}$ (see Section 3), and let $\hat D y(x) = 0$ be a linear 
second-order ODE with non-constant coefficients.  The form of functions $B(x)$ 
and $C(x)$ can be selected so that the resulting equations represent all the 
members of $\mathcal{Q}_{sf}$.  For a special case of $B(x) = 0$, we~define 
$\hat D_o = {{d^2} / {dx^2}} + C(x)$, and thereby we~have $\hat D_o y(x) = 0$ with a 
different family the SF solutions.  In~general, the solutions of the ODEs of 
$\mathcal{Q}_{sf}$ can be written in the following form $y(x) = c_1 y_1(x) 
+ c_2 y_2(x)$, where~$c_1$ and $c_2$ are integration constants [1,2,], and 
$y_1 (x)$ and $y_2(x)$ are given in terms of the SFs; the~solutions written 
in this form are used in Section 2 of this paper.  

Typically, the ODEs of $\mathcal{Q}_{sf}$ are obtained by separation of variables 
in hyperbolic, parabolic and elliptic partial differential equations (PDEs) [1,2,3].  Another 
(less known) method is based on Lie groups, whose irreducible representations (irreps) 
are used to find the SF and their corresponding ODEs~[1,6].  There have also been 
some attempts to establish the Lagrangian formalism for the ODEs of $\mathcal{Q}_{sf}$ 
(e.g., [7,8,9]); however, so far, the problem has not yet been fully solved for $\hat D y(x) = 0$.  
Therefore, the main objective of this paper is to establish the Lagrangian formalism for 
the ODEs of $\mathcal{Q}_{sf}$ and derive new Lagrangians for these equations.

The existence of Lagrangians is guaranteed by the Helmholtz conditions [10], which 
can also be used to derive the Lagrangians.  In general, the Helmholtz conditions 
allowed for the existence of Lagrangians for the ODEs of the form $\hat D_o y(x)
= 0$, but they prevent the Lagrangian formalism for $\hat D y(x) = 0$ because of 
the presence of the term with the first order derivative (e.g., [11,7]).  The~procedure 
of finding the Lagrangians is called the inverse (or Helmholtz) problem of the calculus 
of variations and there are different methods to solve this problem (e.g., [12,13,14,15]).  
In this paper, the~Helmholtz problem is solved differently and new Lagrangians for 
the ODEs of $\mathcal{Q}_{sf}$ are derived.  A~special emphasis is given to the 
validity of the Helmholtz conditions for the derived Lagrangians.  We~also explore 
applications of the obtained results to the Airy, Bessel, Legendre and Hermite 
equations.

There are two main families of Lagrangians, the so-called {\it standard} 
and {\it non-standard} Lagrangians.  The standard Lagrangians (SLs) are 
typically expressed as the difference between terms that can be identified 
as the kinetic and potential energy [14].  A broad range of different methods 
exists, and~these methods were developed to obtain the SLs  for both linear 
and non-linear ODEs, and PDEs.   Some methods involve the concept 
of the Jacobi last multiplier [16,17,18] or use fractional derivatives [19], and 
others are based on different transformations that allow deriving the SLs  
for the conservative and non-conservative physical systems described by 
either linear or non-linear ODEs [20].  There are also methods for finding 
the SLs  for linear second-order PDEs, including the wave, Laplace and 
Tricomi-like equations [21].   

A procedure of deriving the SLs  may also give the Lagrangians for which 
the Euler--Lagrange (E--L)
equation vanishes identically [22].  These Lagrangians may have 
terms that depend on both the dependent variable and its first derivative
[22], and terms that resemble the potential energy term; thus,
we call such Lagrangians the mixed Lagrangians (MLs).   In general, 
for the ODEs considered in this paper, these Lagrangians are not null 
Lagrangians (NLs), as the latter also require that they can be expressed 
as the total derivative of a scalar function [22,23,24], which is often called 
the gauge function [25].  However, there are some interesting exceptions 
from this rule and they are explored in this paper.  The derived MLs are 
new; however, the obtained SLs are already known (e.g., [10,11,24]) 
and they appear here only as a byproduct of the procedure that is used.  

The non-standard (or not-natural) Lagrangians (NSLs ) are such that 
identification of the kinetic and potential energy terms cannot
be done, and therefore these NSLs  are simply the generating functions
for the original equations, as first pointed out by Arnold [15].  There have 
been many attempts to obtain the NSLs  for different ODEs.  One of the 
first application of the NSLs  to physics was done by Alekseev and Arbuzov 
[26], who formulated the Yang--Mills field theory using NSLs , and thus
demonstrated the usefulness of the NSLs  for the fundamental theories 
of modern physics.  Different forms of the NSLs  have been proposed 
and applied to different physical problems~[8,9,27,28], including a new 
NSL introduced by El-Nabulsi [29], who published a number of papers 
with interesting applications ranging from quantum fields and particle 
physics to general relativity and cosmology.  Moreover, some NSLs  
were used by other authors, who established the Lagrangian formalism 
for Riccati [18] and Lienard [30] equations.  Here, we present only one 
special family of NSLs and we refer to these Lagrangians as the NSLs 
throughout the paper.  

Despite the efforts described above, the NSLs  do not have yet a 
well-established space in the theory of inverse variational problems.
In our recent work [31], we studied the generalised NSLs and 
demonstrated that finding such Lagrangians requires solving a 
Riccati equation, whose solution introduces a new dependent variable.
The E--L equation forces this variable to appear in the original equation, 
and our results show that in order to remove it from this equation, the 
Lagrangian formalism must be amended by an auxiliary condition, 
which is a novel phenomenon in the calculus of variations.  By using 
the condition, the new variable is removed and the original equation 
is obtained.  In this paper, we investigate the phenomenon in detail and 
present the auxiliary conditions for the ODEs of $\mathcal{Q}_{sf}$.

The outline of the paper is as follows: in Section 2, the Lagrangian formalism 
for the ODEs of $\mathcal{Q}_{sf}$ is established using both standard and 
non-standard Lagrangians, and validity of the Helmholtz conditions is also 
explored; in Section 3, applications of the obtained results to some selected 
ODEs of $\mathcal{Q}_{sf}$ are given and discussed; and our concluding 
remarks and open problems are presented in Section 4.

\section{Lagrangian Formalism}

\subsection{Hamilton's Principle and the Existence of Lagrangians}

From a mathematical point of view, in the Lagrange formalism we are 
provided with a functional $\mathcal{S} [y(x)]$, which depends on 
an ordinary and smooth function $y(x)$ that must be determined.  
The~functional is a map $\mathcal{S} :\ \mathcal{C}^{\infty}(\mathcal{R}) 
\rightarrow \mathcal{R}$, with $\mathcal{C}^{\infty}(\mathcal{R})$ 
being a space of smooth functions.  Typically $\mathcal{S} [y(x)]$ 
is defined by an integral over a smooth function $L$ that depends 
on $y'(x) =  dy / dx$, $y$ and on $x$, and the function $L (y', y, x)$ 
is called the Lagrangian function or simply Lagrangian.  The functional 
$\mathcal{S} [y(x)]$ defined in this way is known as the functional 
action, or simple action, and the principle of least action, or Hamilton's 
principle [14], requires that $\delta \mathcal{S} = 0$, where $\delta$ 
is the variation defined as the functional (Fr\'echet) derivative of 
$\mathcal{S} [y(x)]$ with respect to $y(x)$.  Using $\delta \mathcal{S} 
= 0$, the Euler--Lagrange (E--L) equation is obtained
\begin{equation}
{d \over {dx}} \left ( {{\partial L} \over {\partial y'}} \right ) - 
{{\partial L} \over {\partial y}} = 0\ ,
\label{S1eq0}
\end{equation}
and this equation becomes a necessary condition for the action to be 
stationary (to have either a minimum or maximum or saddle point).  

In general, the E--L equation leads to a second-order ODE that can be 
further solved to obtain $y(x)$ that makes the action stationary.  The
described procedure is the basis of the classical calculus of variations, 
and it works well when the Lagrangian $L (y', y, x)$ is already given.  
Deriving the second-order ODE from the E--L equation is known as the 
{\it Lagrangian formalism}, and in this paper we deal exclusively with 
this formalism.  Our main goal is to establish the Lagrangian formalism 
for the ODEs of  ${\mathcal{Q}_{sf}}$ and find new Lagrangians.  

To fully establish the Lagrangian formalism for the ODEs of 
$\mathcal{Q}_{sf}$, we must know how to construct the SLs, NLs 
and NSLs  for these equations.  In the following, we describe new 
families of Lagrangians and show that the existence of some of 
these Lagrangians has profound implications on the validity of 
the Helmholtz conditions and on the calculus of variations.  

\subsection{Standard and Mixed Lagrangians}

In the previous work [8], a very effective method of finding some SLs  
for $\hat D y(x) = 0$, which~includes the ODEs of $\mathcal{Q}_{sf}$,
was proposed.  The constructed SLs, denoted here as $L_s$ are 
of the following~form:
\begin{equation}
L_s [y'(x), y(x), x] = G_s [y'(x), y(x), x]\ E_{s} (x)\ ,
\label{S1eq1}
\end{equation}
where 
\begin{equation}
G_s [y'(x), y(x), x] = {1 \over 2} \left [ \left ( y^{\prime} (x) 
\right )^2 - C(x) y^2 (x) \right ]\ , 
\label{S1eq2}
\end{equation}
and $E_{s} (x) = \exp{ [ \int^x B(\tilde x) d \tilde x ]}$.  Since 
$E_{s} (x)$ is only a function of one independent variable $x$, 
the lower limit must be an arbitrary constant, which can be omitted
because such constant has no effect on the Lagrangian formulation.  
Note that in a special case of $B (x)$ = constant,
$L_{s} [y'(x), y(x), x]$ 
becomes the Caldirola--Kanai Lagrangian [32,33].  

As shown in [9], the equation $\hat D y(x) = 0$ can also be derived 
from the following Lagrangian   
\begin{equation}
L_{sm} [y'(x), y(x), x] = L_s [y'(x), y(x), x] + L_m [y'(x), y(x), x]\ ,
\label{S1eq3}
\end{equation}
where 
\begin{equation}
L_m [y'(x), y(x), x] = G_{m} [y'(x), y(x), x]\ E_s (x)\ ,
\label{S1eq4}
\end{equation}
and 
\begin{equation}
G_m [y'(x), y(x), x] = {1 \over 2} B(x) y (x) y^{\prime}(x) 
+ {1 \over 4} \left [ B^{\prime} (x) + B^2 (x) \right ] y^2 (x)\ .
\label{S1eq5}
\end{equation}

Since in $L_m [y'(x), y(x), x]$ the variables $y^{\prime} (x)$ and 
$y (x)$ are mixed when compared to $L_s [y'(x), y(x), x]$, we 
call $L_m [y'(x), y(x), x]$ the mixed Lagrangian to distinguish it
from the standard Lagrangian.  

Having defined the Lagrangians $L_{sm}$, $L_s$ and $L_m$, we 
now state our main result concerning these Lagrangians in the 
following proposition.

{\bf Proposition:}
Let $L_s$ and $L_{sm}$ be the Lagrangians 
given respectively by Equations (\ref{S1eq1}) and (\ref{S1eq3}), and let $L_m$ 
be defined by Equation (\ref{S1eq4}).   Both $L_s$ or $L_{sm}$ give the same 
$\hat D y(x) = 0$ if, and only if, $L_m$ makes the E--L equation vanish 
identically.

{\bf Proof:}
The proof is straightforward, as substitution of $L_s 
[y'(x) , y(x), x]$ or $L_{sm} [y'(x), y(x), x]$ into the E--L equation 
gives the same original equation $\hat D y(x) = 0$ because 
\begin{equation}
{d \over {dx}} \left ( {{\partial L_m} \over {\partial y'}} \right ) 
= {1 \over 2} \left [ B'(x) y (x) + B(x) y^{\prime}(x) + B^2 (x) y (x)
\right ] E_s (x)\ ,
\label{S1eq6}
\end{equation}
and   
\begin{equation}
{{\partial L_m} \over {\partial y}} = {1 \over 2} \left [ B'(x) y (x) 
+ B(x) y^{\prime}(x) + B^2 (x) y (x) \right ] E_s (x)\ ,
\label{S1eq7}
\end{equation}
are exactly equal, which means that they make the E--L equation 
(see Equation (\ref{S1eq0})) vanish identically.  
\begin{equation}
{d \over {dx}} \left ( {{\partial L_m} \over {\partial y'}} \right ) 
= {{\partial L_m} \over {\partial y}}\ .
\label{S1eq8}
\end{equation}

As a result, $L_m$ makes null contributions to the Lagrange formalism, 
and therefore has no influence on derivation of the original equation 
$\hat D y(x) = 0$.  This concludes the proof.

Important results that are consequences of Proposition 1 are now 
presented in the following three corollaries, which require the following 
definition:  a Lagrangian is called the {\it simplest} if, and only if, this 
Lagrangian does not contain $L_m [y'(x), y(x), x]$.

{\bf Corollary:}
The Lagrangian $L_s [y'(x), y(x), x]$ given 
by Equation (\ref{S1eq1}) is the simplest standard Lagrangian for the 
ODEs of $\mathcal{Q}_{sf}$.

{\bf Corollary:}
The Lagrangian $L_m [y'(x), y(x), x]$ given 
by Equation (\ref{S1eq4}) can be added to any known Lagrangian without 
making any changes in the resulting original equation.

{\bf Corollary:}
The Lagrangian $L_m [y'(x), y(x), x]$, with its 
different coefficients $B(x)$ for different ODEs of ${\mathcal{Q}_{sf}}$, 
forms a new family of Lagrangians that make the E--L equation 
vanish identically.

\subsection{Null Lagrangian  and Gauge Functions}

Our results presented above show that for $L_m [y'(x), y(x), x]$, the 
E--L equation vanishes identically, which means that the mixed 
Lagrangians have null effects on the Lagrangian formalism.  The 
mixed Lagrangian may or may not become the null Lagrangian 
$L_n [y'(x), y(x), x]$ that is defined [22] as
\begin{equation}
L_n [y'(x), y(x), x] = {{d \phi} \over {d x}}\ ,
\label{S1eq9}
\end{equation}
where $\phi (x)$ is a gauge function [25].  It is easy to verify that for 
the ODEs of ${\mathcal{Q}_{sf}}$, such a function cannot be uniquely 
defined, except some special cases considered below and also in 
Section 3.   Thus, in general $L_m [y'(x), y(x), x] \neq L_n [y'(x), 
y(x), x]$ for most ODEs of $\mathcal{Q}_{sf}$; however, there 
are a few special cases when the MLs are the NLs.  Here is one 
interesting example.

The ODEs of ${\mathcal{Q}_{sf}}$ reduce to $\hat D_o y (x) = 
0$ if $B (x) = 0$.  In this case,  $L_m [y'(x), y(x), x] = 0$ but
we may consider $L_{mo} [y'(x), y(x)] = q y^{\prime} (x) y (x) 
/ 2$, with $q =$ const, and show that this is the mixed Lagrangian 
for $D_o y (x) = 0$.  In addition, it is easy to demonstrate the 
existence of the gauge function because  
\begin{equation}
L_{mo} [y'(x), y(x), x] = {{d \phi} \over {d x}}\ ,
\label{S1eq10a}
\end{equation}
which means that for these ODEs the derived mixed Lagrangians 
are the null Lagrangians and that the gauge function can be derived.
The resulting gauge function is given by 
\begin{equation}
\phi (x) = {1 \over 8} q y^2 (x)\ .
\label{S1eq10b}
\end{equation}

Since $q$ is arbitrary, our results show that $\phi (x)$ is the 
gauge function for the ODEs of the form $\hat D_o y (x) = 0$ (see 
Section 3 for applications).   

Let us now assume that $q \neq$ const and that $q (x) = B (x)$.
By making this assumption, we want to show that this does not 
cause $\phi (x)$ to be the gauge function for the ODEs given by 
$\hat D y (x) = 0$ because one extra term must be added, and 
this term cannot be included into the total derivative.  Therefore, 
for most ODEs of ${\mathcal{Q}_{sf}}$, their $L_m [y'(x), y(x), 
x]$ are not the same as $L_n [y'(x), y(x), x]$ .

Another important point is that the obtained NLs can be easily 
eliminated (e.g., [25]).  Nevertheless, our purpose of deriving 
the NLs is motivated by some previous work [22,23,24] in which 
it was clearly shown that the NLs are very useful for identifying 
symmetries in physical systems, and that they also play a 
significant role in Carath\'eodory's theory of fields of extremals 
and integral invariants [23].  Moreover, the gauge functions 
resulting from the derived NLs may have some effects on the 
behaviours of quantum systems whose solutions are known to 
be given by the SFs of mathematical physics.   

The procedure of eliminating the NLs from the Lagrangians is 
based on the fact that these NLs are expressed as the total 
derivative of an arbitrary scalar function [25].  Therefore, 
the same elimination procedure cannot be applied to the 
MLs, whose presence may actually give a different view 
of symmetries in physical systems.

\subsection{Non-Standard Lagrangians}

Having established the Lagrangian formalism based on the SLs  
and deriving the corresponding NLs, we now develop the Lagrangian 
formalism based on the NSLs.  Our new results are presented by
 the following two propositions.

{\bf Proposition:}
Let $L_{ns} [y'(x) , y(x), x] = 1 / [f (x) y^{\prime} 
(x) + g (x) y (x)]$ be a non-standard Lagrangian with $f (x)$ and $g (x)$ 
being ordinary and smooth functions.  The Lagrangian formalism can be 
used to determine these functions for any ODE of $\mathcal{Q}_{sf}$, 
and expressing $L_{ns} [y'(x) , y(x), x]$ in the following form
\begin{equation}
L_{ns} [y^{\prime} (x), y(x), x] = H_{ns} [y^{\prime} (x), y(x), x]\ 
E_{ns} (x)\ ,
\label{S1eq11}
\end{equation}
where
\begin{equation}
H_{ns} [y^{\prime} (x), y(x), x] = {{1} \over {[y^{\prime} (x) \bar v (x) 
- y (x) \bar v^{\prime} (x)]\ \bar v^2 (x)}}   
\label{S1eq12}
\end{equation}
\noindent
and $E_{ns} (x) = \exp{[ - 2 \int^x B(\tilde x) d \tilde x]}$, with the 
necessary auxiliary condition $\hat D \bar v (x)$ $= 0$.

{\bf Proof:}
Substituting $L_{ns} [y'(x) , y(x), x] = {{1} / [{f (x) 
y^{\prime} (x) + g (x) y(x)}}]$ into the Euler--Lagrange equations, 
we obtain 
\begin{equation}
{{d} \over {dx}} \left ( {{\partial L_{ns}} \over {\partial y^{\prime}}}
\right ) = - {{f^{\prime}} \over {(f y^{\prime} + g y )^2}} + 2 
{{(f^{\prime} y^{\prime} + f y^{\prime \prime} + g^{\prime} y 
+ g y^{\prime} ) f} \over {(f y^{\prime} + g y )^3}}\ ,
\label{S1eq13a}
\end{equation}
and 
\begin{equation}
\left ( {{\partial L_{ns}} \over {\partial y}} \right ) = - {g \over 
{(f y^{\prime} + g y )^2}}
\label{S1eq13b}
\end{equation}
which gives
\begin{equation}
y^{\prime \prime} + {1 \over 2} \left ( {{f^{\prime}} \over {f}} + 
{{3g} \over {f}} \right ) y^{\prime} + \left ( {{g^{\prime}} 
\over {f}} - {{f^{\prime} g} \over {2f^2}} + {{g^2} \over 
{2f^2}} \right ) y = 0\ . 
\label{S1eq13}
\end{equation}

By comparing this equation to $\hat D y(x) = 0$, the following two 
conditions that allow finding $f(x)$ and $g(x)$ are obtained
\begin{equation}
{1 \over 2} {{f^{\prime}} \over {f}} + {3 \over 2} {{g} \over 
{f}} = B(x)\ , 
\label{S1eq14}
\end{equation}
and
\begin{equation}
{{g^{\prime}} \over {f}} - {1 \over 2} {{f^{\prime} g} \over 
{f^2}} + {1 \over 2} {{g^2} \over {f^2}} = C(x)\ , 
\label{S1eq15}
\end{equation}
\noindent
with $f(x) \neq 0$.  From Equation (\ref{S1eq14}), we get 
\begin{equation}
{{g (x)} \over {f (x)}} = {2 \over 3} B (x) - {1 \over 3} 
{{f^{\prime} (x)} \over {f (x)}}\ .
\label{S1eq16a}
\end{equation}
and 
\begin{equation}
{{g^{\prime} (x)} \over {f (x)}} = {2 \over 3} B^{\prime} (x) - 
{1 \over 3} {{f^{\prime} (x)} \over {f (x)}} + {2 \over 3} B (x) 
{{f^{\prime} (x)} \over {f (x)}}  - {1 \over 3} \left [ {{f^{\prime} 
(x)} \over {f (x)}} \right ]\ .
\label{S1eq16b}
\end{equation}

Substituting Equations (\ref{S1eq16a}) and (\ref{S1eq16b})
into Equation (\ref{S1eq15}), we find 
\begin{equation}
{2 \over 3} \left [ B^{\prime} (x) + {1 \over 3} B^2 (x) \right ]
- {1 \over 3} {{f^{\prime} (x)} \over {f (x)}} - {1 \over 9} 
\left [ {{f^{\prime} (x)} \over {f (x)}} \right ]^2 + {1 \over 9}
{{f^{\prime} (x)} \over {f (x)}} B (x) = C (x)\ .
\label{S1eq16c}
\end{equation}

Introducing $u(x) = f^{\prime}(x) / f(x)$, the following Riccati 
equation for $u(x)$ is obtained
\begin{equation}
u^{\prime}   + {1 \over 3} u  ^2 - {1 \over 3} u   B (x) - \left [ 
{2 \over 3} B^2 (x) + 2 B'(x) - 3 C (x) \right ] = 0\ . 
\label{S1eq16}
\end{equation}

With $f(x)\ =\ \exp {\left [ \int^{x} u (\tilde x)\ d \tilde x \right ]}$ 
and $g (x)\ =\ 2 \left [ B(x) - u (x) / 2 \right ] f (x) / 3$, it is seen 
that finding $u (x)$, which satisfies the Ricatti equation, allows us 
to determine the functions $f (x)$ and $g (x)$.

We now transform Equation (\ref{S1eq16}) by introducing a new variable 
$v (x)$, which is related to $u (x)$ by $u (x) = 3 v^{\prime} (x) 
/ v (x)$ with $v(x) \neq 0$, and obtain
\begin{equation}
v^{\prime \prime} + B(x) v^{\prime} + C(x) v = F (v^{\prime}, 
v, x)\ ,
\label{S1eq17}
\end{equation}
where $F (v^{\prime}, v, x) = 2\ [ 2 B(x) v^{\prime} + B'(x) 
v + B^2 (x) v / 3 ] / 3$.  

Let us now transform Equation (\ref{S1eq17}) by using 
\begin{equation}
v (x) = \bar v (x)\ \exp{\left [ \int^x \chi(\tilde x) d \tilde x 
\right ]}\ ,
\label{S1eq19}
\end{equation}
and obtain 
\begin{equation}
\bar v^{\prime \prime} (x) + B(x) \bar v^{\prime} (x) + 
C(x) \bar v (x) = 0\ ,
\label{S1eq18}
\end{equation}
if, and only if, $\chi(x) = - 2 B(x) / 3$.  This allows writing the 
solution for $u (x)$ in the following form
\begin{equation}
u (x) = 3 {{\bar v^{\prime} (x)} \over {\bar v (x)}} + 2 B(x)\ .
\label{S1eq20}
\end{equation}

It is easy to verify that Equation (\ref{S1eq20}) is the solution of the 
Riccati equation given by Equation~(\ref{S1eq16}), if Equation (\ref{S1eq18}) 
is taken into account.   Having obtained $u (x)$, the functions $f (x)$ 
and $g (x)$ can be calculated
\begin{equation}
f (x) = \bar v^3 (x)\ \exp{\left [ 2 \int^x B(\tilde x) 
d \tilde x\right ]}\ ,
\label{S1eq21}
\end{equation}
and 
\begin{equation}
g (x) = - {{\bar v^{\prime} (x)} \over {\bar v (x)}} \bar v^3 (x)\ 
\exp{\left [ 2 \int^x B(\tilde x) d \tilde x \right ]}\ ,
\label{S1eq22}
\end{equation}
and the resulting non-standard Lagrangian is given by 
\begin{equation}
L_{ns} [y^{\prime} (x), y(x), x] = H_{ns} [y^{\prime} (x), y(x), x]   
E_{ns} (x)\, 
\label{S1eq23}
\end{equation}
where
\begin{equation}
H_{ns} [y^{\prime} (x), y(x), x] = {{1} \over {[y^{\prime} (x) \bar v (x) 
- y (x) \bar v^{\prime} (x)]\ \bar v^2 (x)}}\ ,   
\label{S1eq24}
\end{equation}
which is the same as that given by Equations (\ref{S1eq13}) and (\ref{S1eq14}).
Since the derived non-standard Lagrangian depends explicitly on $\bar v 
(x)$, the auxiliary condition $\hat D \bar v (x) = 0$ (see Equation (\ref{S1eq18})) 
must supplement $L_{ns} [y^{\prime} (x), y(x), x]$.   This concludes 
the proof.

Having derived the NSLs  for the ODEs of $\mathcal{Q}_{sf}$, we must 
now verify that the original ODEs can be obtained from the non-standard 
Lagrangian and its auxiliary condition.  The following proposition and 
corollaries present our results.

{\bf Proposition:}
The Lagrange formalism based on the non-standard 
Lagrangians can be established for the ODEs of $\mathcal{Q}_{sf}$ if, and 
only if, $L_{ns} [y^{\prime} (x), y(x), x]$ of Proposition 2 is used together 
with the auxiliary condition $\hat D \bar v(x) = 0$.

{\bf Proof:}
Substituting the definition of $L_{ns} [y^{\prime}(x), y(x), x]$ 
given by Equations (\ref{S1eq11}) and (\ref{S1eq12}) into the E--L equation, we 
obtain  
\begin{equation}
{{y^{\prime \prime} (x) \bar v (x) - y (x) \bar v^{\prime \prime} 
(x)} \over {y^{\prime} (x) \bar v (x) - y (x) \bar v^{\prime} (x)}} 
+ B(x) = 0\ ,
\label{S1eq25}
\end{equation}
which can also be written as
\begin{equation}
\left [ y^{\prime \prime} (x) + B(x) y^{\prime} (x) \right ] 
\bar v (x) = \left [ \bar v^{\prime \prime} (x) + B(x) \bar 
v^{\prime} (x) \right ] y(x)\ .
\label{S1eq26}
\end{equation}

This is the result of substituting $L_{ns} [y^{\prime} (x), 
y (x), x]$ obtained in Proposition 1 into the E--L equation,  
and it is seen that Equation (\ref{S1eq26}) is not the same as the 
original equation $\hat D y(x) = 0$.  In order to derive the 
original equation, we must now use the auxiliary condition 
(see Equation (\ref{S1eq18})) that gives
\begin{equation}
\bar v^{\prime \prime} (x) + B(x) \bar v^{\prime} (x) = 
- C(x) \bar v (x)\ .
\label{S1eq27}
\end{equation}

This shows that by applying the the auxiliary condition 
$\hat D \bar v (x) = 0$, the E--L equation gives the 
original equation $\hat D y(x) = 0$, which concludes 
the proof.

{\bf Corollary:}
The Lagrangian $L_{ns} [y'(x), y(x), x]$ 
given by Equation (\ref{S1eq11}) is the non-standard Lagrangian for 
the ODEs of $\mathcal{Q}_{sf}$.

{\bf Corollary:}
All non-standard Lagrangians $L_{ns} [y'(x), 
y(x), x]$ given by Equation (\ref{S1eq11}) form a new and separate family 
among all known non-standard Lagrangians.

Since the solutions of the ODEs of $\mathcal{Q}_{sf}$ are given by 
the SFs, the same functions are the solutions for the auxiliary condition 
$\hat D \bar v (x) = 0$.  Nevertheless, the dependent variables for 
which the solutions are known are not the same, and therefore the
integration constants must be different to obey different boundary 
conditions the two variables satisfy.  The implications of this are 
shown by the following~corollary.

{\bf Corollary:}
Let $y(x) = c_1 y_1(x) + c_2 y_2(x)$
and $\bar v(x) = \bar c_1 y_1(x) + \bar c_2 y_2(x)$ be the
superpositions of linearly independent solutions of $\hat D y (x) 
= 0$ and $\hat D \bar v (x) = 0$, respectively, with $c_1$, 
$c_2$, $\bar c_1$ and $\bar c_2$ being the integration 
constants.  Then, the function $H_{ns} [y^{\prime} (x),
 y(x), x]$ of Proposition 2 becomes
\[
H_{ns} [y^{\prime}_1 (x), y^{\prime}_2 (x), y_1 (x), y_2 (x), x] 
\]
\begin{equation}
\hskip0.25in = (c_1 \bar c_2 - \bar c_1 c_2)^{-1}\ 
[y^{\prime}_1(x) y_2 (x) - y_1(x) y^{\prime}_2 (x)]^{-1}\ 
[\bar c_1 y_1(x) + \bar c_2 y_2(x)]^{-2}\ ,
\label{S1eq28}
\end{equation}
where $c_1 \bar c_2 \neq \bar c_1 c_2$; the term $[y^{\prime}_1(x) 
y_2 (x) - y_1(x) y^{\prime}_2 (x)]$ is the non-zero Wronskian; and 
the term $[\bar c_1 y_1(x) + \bar c_2 y_2(x)]^2$ is also non-zero 
for both oscillatory and non-oscillatory solutions.

Let us point out that the solutions $y_1(x)$ and $y_2(x)$ are the 
same for both $y(x)$ and $\bar v(x)$, which~means that their 
dependence on $x$ is identical.  However, the integration constants 
are different because $y(x)$ and $\bar v(x)$ are not the same 
variables, and therefore they must obey different boundary conditions. 

The requirement of Corollary 6 that $[\bar c_1 y_1(x) + \bar c_2 
y_2(x)]^2 \neq 0$ needs additional explanation.  Clearly, the 
requirement is non-zero when both $y_1(x)$ and $y_2(x)$ are 
non-oscillatory.  Moreover, if both $y_1(x)$ and $y_2(x)$ are 
oscillatory, then the requirement still remains non-zero because
the locations of the zeros of the two linearly independent solutions 
given by the SFs are never the same.  The exception could be the
point $x = 0$; thus, we consider $x \in (0, \infty)$ in our 
applications (see Section \ref{sec3}). 

A novel result is that the original ODEs have only one dependent 
variable and that the procedure of deriving the NSLs  introduces a
new dependent variable $\bar v (x)$, which explicitly appears in 
the NSLs.  Let us point out that these NSLs  do give the original
ODEs but only with the auxiliary condition that allows eliminating
the additional dependent variable  [31].   

To show that the auxiliary condition is necessary, we substitute
the NSLs  given by Equation (\ref{S1eq23}) into the E--L equation written 
for the variable $\bar v (x)$, and obtain  
\begin{equation}
{{y^{\prime \prime} (x) \bar v (x) - y (x) \bar v^{\prime \prime} 
(x)} \over {y^{\prime} (x) \bar v (x) - y (x) \bar v^{\prime} (x)}} 
+ B(x) = \left [ {{\bar v^{\prime} (x)} \over {\bar v (x)}} \right ]
- 2 \left [ {{y^{\prime} (x)} \over {y (x)}} \right ]\ .
\label{S1eq29}
\end{equation}

Comparing this result to Equation (\ref{S1eq25}), it is seen that 
there are two extra terms on the right-hand-side (RHS)
of Equation (\ref{S1eq29}), which can be 
written as   
\[
\left [ y^{\prime \prime} (x) + B(x) y^{\prime} (x) \right ] 
\bar v (x) = \left [ \bar v^{\prime \prime} (x) + B(x) \bar 
v^{\prime} (x) \right ] y(x)  \left [ {{\bar v^{\prime} (x)} 
\over {\bar v (x)}} \right ] 
\]
\begin{equation}
\hskip0.50in \times  \left [ y^{\prime} (x) \bar v (x)
- y (x) \bar v^{\prime} (x) \right ]\ ,
\label{S1eq30}
\end{equation}
which shows that the additional term on the RHS of this 
equation can only be eliminated when $y^{\prime} (x) 
\bar v (x) - y (x) \bar v^{\prime} (x) = 0$.  However, 
this violates the main result of Corollary 6 that the 
Wronskian must be non-zero.  

Thus, our results clearly demonstrate that the auxiliary condition 
cannot be derived from the E--L equation, but instead it must be 
obtained independently (see Proposition 3).  Only with this 
independently derived auxiliary condition, can the original ODEs 
be obtained from the NSLs; this is a new phenomenon 
in the calculus of variations, as we already pointed out in [31].

\subsection{Helmholtz Conditions and Their Validity}

The existence of Lagrangians is guaranteed by the Helmholtz conditions
(HCs), which are necessary and sufficient conditions [10,27].  Let 
$F_i (y_j^{\prime \prime}, y_j^{\prime}, y_j, x) = 0$ be a set of $n$ 
ODEs, with $i = 1, 2, ..., n$ and $j = 1, 2, ..., n$;
then, the Helmholtz conditions are 
\begin{equation}
{{\partial F_i} \over {\partial y_j^{\prime \prime}}} = 
{{\partial F_j} \over {\partial y_i^{\prime \prime}}}\ , 
\label{S1eq31a}
\end{equation}
\begin{equation}
{{\partial F_i} \over {\partial y_j}} - {{\partial F_j} \over {\partial y_i}} 
= {1 \over 2} {d \over {dx}} \left ( {{\partial F_i} \over {\partial 
y_j^{\prime}}} - {{\partial F_j} \over {\partial y_i^{\prime}}} \right )\ ,
\label{S1eq31b}
\end{equation}
and
\begin{equation}
{{\partial F_i} \over {\partial y_j^{\prime}}} + {{\partial F_j} \over 
{\partial y_i^{\prime}}} = 2 {d \over {dx}} \left ( {{\partial F_j} \over 
{\partial y_i^{\prime \prime}}} \right )\ .
\label{S1eq31c}
\end{equation}

Since for the ODEs of $\mathcal{Q}_{sf}$, $i = j = 1$, the first and 
second conditions are trivially satisfied; however, the third HC is not 
satisfied.  The reason is that the LHS = $B(x)$ but the RHS = 0; thus, 
the third HC fails to be valid, and this implies that no Lagrangian can 
be constructed for any ODEs of $\mathcal{Q}_{sf}$ with $B(x) \neq 0$.  
Despite this strong negative statement, the Lagrangians obtained in this 
paper (as well as in some previous papers, notable in [5]) seem to 
contradict the third HC. 
 
Let us explain the contradiction by substituting $L_s$ given by Equation 
(\ref{S1eq1}) into the E--L equation.  The result is 

\begin{equation}
\left [ y^{\prime \prime} + B(x) y^{\prime} + C(x) y 
\right ] E_s (x) = 0\ , 
\label{S1eq32}
\end{equation}
and it is easy to verify that this equation does satisfy the third HC and 
also the first and second HCs; therefore, the existence of $L_s$ is justified 
by the HCs.  However, the problem is that Equation (\ref{S1eq32}) is {\it not} 
the same as the original equation ($\hat D y (x) = 0$), and it does not 
even belong to $\mathcal{Q}_{sf}$.  In other words, the~Lagrangian $L_s$ 
is consistent with the HCs but it gives the equation that differs from the 
original one by the factor $E_s (x)$.  Since $E_s (x) > 0$, $[\hat D y (x)] 
E_s (x) = 0$ can be divided by $E_s (x)$ in order to obtain the original 
equation.  This shows that the contradiction arises because the HCs do 
not account for the required division by $E_s (x)$.  The credit for 
discovering and explaining this phenomenon goes back to Bateman [5].  

Surprisingly, for the non-standard Lagrangians, the results are 
different than those obtained above for the standard Lagrangians.
This can be shown by substituting $L_{ns}$ given by Equations (\ref{S1eq11}) 
and (\ref{S1eq12}) into the E--L equation, and finding
\begin{equation}
\left [ y^{\prime \prime} + B(x) y^{\prime} + C(x) y 
\right ] \bar v (x) E_{ns} (x) = 0\ , 
\label{S1eq33}
\end{equation}
where $\bar v (x)$ is a solution to the auxiliary condition, which is
introduced in Proposition 2 together with $E_{ns} (x)$.  Clearly, 
this equation does not satisfy the third HC because of the presence
of $E_{ns} (x)$.  

Nevertheless, taking $\bar v (x) = \bar v_o [E_s (x)]^3$, where 
$\bar v_o$ is an integration constant, enables making Equation (\ref{S1eq33}) 
the same as Equation (\ref{S1eq32}).  This is important because Equation 
(\ref{S1eq32}) already satisfies the third HC.  The only problem that 
remains to be resolved is whether the imposed solution on $\bar v (x)$ 
is also a solution of the auxiliary condition.  In general, this will not be
the case; however, the ODEs whose coefficients $B (x)$ and $C (x)$
are related by $B^{\prime} (x) + 4 B^2 (x) = - C(x) / 3$ will be the
exception.  Thus, the presented results show that there is only a small 
subset of $\mathcal{Q}_{sf}$ for which the existence of the NSLs  can 
be justified by the HCs.  A new phenomenon is that the remaining ODEs 
have their non-standard Lagrangians despite the fact that the resulting 
ODEs violate the HCs.

Let us point out that the problems described above for $\hat D y (x) = 0$ 
do not exist for $\hat D_o y (x) = 0$, since for the latter the coefficient 
$B (x) = 0$, and as a result, the Helmholtz conditions are satisfied.  It is 
also important to emphasise that the existence of the MLs and NLs is 
independent from the Helmholtz conditions, because these Lagrangians 
have no effects on the derivation of the original ODEs.  In other words, 
once the mixed or null Lagrangian are found, they can be added to 
any standard or non-standard Lagrangian without changing the original 
equation or affecting the HCs.

\subsection{Implications of Our Results on Calculus of Variations}

We formally established the Lagrangian formalism for the ODEs 
of $\mathcal{Q}_{sf}$, and demonstrated that this can be achieved 
by using either standard or non-standard Lagrangians.  We also 
derived the mixed Lagrangians and identified the ODEs for which
these Lagrangians become the null Lagrangians.  Knowing the NLs, 
we obtained their corresponding gauge functions and discussed
their role in the Lagrangian formalism.  Our results have profound 
implications on the calculus of variations.  

First, we showed that the standard Lagrangians, previously derived
for the ODEs of $\mathcal{Q}_{sf}$, may~have their corresponding 
mixed Lagrangians, which cannot be used to obtain the original 
ODEs.  We~demonstrated that the mixed Lagrangians can be 
determined for all considered ODEs but the null Lagrangians exist 
only for some special cases (including the ODEs of the form $\hat 
D_o y (x) = 0$), and that only in these cases the corresponding 
gauge functions can be derived.  The role of the null Lagrangians 
in studies of symmetries of physical systems and other phenomena 
was also briefly discussed.   It was pointed out that the MLs may 
also contribute to these studies.   

Second, in order to obtain the NSLs for the ODEs of $\mathcal{Q}_{sf}$, 
we had to solve the non-linear Riccati equation (see Equation (\ref{S1eq18})) 
whose solutions introduced a new dependent variable.  Thus, despite 
the fact that the original ODEs have only one dependent variable,
another one naturally appeared only in the NSLs and not in the SLs .

Third, we demonstrated that this additional dependent variable can 
only be removed by an auxiliary condition, which becomes the 
amendment to the E--L equation.  The existence of the auxiliary 
condition in the Lagrangian formalism based on the NSLs  is a 
new phenomenon in the calculus of variations and it has a 
profound implications on the Helmholtz conditions and their 
validity. 

Fourth, the dependence of the SLs  ($L_s$) and the NSLs  ($L_{ns}$) 
on $B(x)$ and $C(x)$ is significantly different.  Moreover, the form of 
SLs changes for different ODEs; however, the basic form NSLs remains 
practically the same for the ODEs of $\mathcal{Q}_{sf}$.  This is a 
novel property of the NSLs derived here, which has not yet been 
observed in other NSLs previously obtained.

Finally, let us point out that our results clearly showed that 
for the derived NLs there is only a small subset of  ODEs of 
$\mathcal{Q}_{sf}$ with $B(x) \neq 0$ for which the Helmholtz 
conditions are satisfied.  However, other ODEs of $\mathcal{Q}_{sf}$ 
have their non-standard Lagrangians but do not obey the Helmholtz 
conditions.  The effects of this novel discovery may become important 
in establishing the Lagrangian formalism based on the NLs.

\section{Applications to Selected Equations}

\subsection{Airy Equation}

The general form of the ODEs of $\mathcal{Q}_{sf}$ can be 
reduced to $\hat D_o y (x) = 0$ by taking $B (x) = 0$.  Then,
by~specifying $C (x) = - x$, we obtain the following Airy equation 
$y^{\prime \prime} (x) - x y(x) = 0$.  Since $B (x) = 0$, the 
standard Lagrangian (see Proposition 1) is 
\begin{equation}
L_{so} [y^{\prime} (x), y (x), x] = {1 \over 2} \left [ 
\left ( y'(x) \right )^2 + x y^2 (x) \right ]\ ,
\label{S2eq1a}
\end{equation}
and the non-standard Lagrangian (see Proposition 3) can be 
written as 
\begin{equation}
L_{nso} [y^{\prime} (x), y (x), x] = H_{nso} [y^{\prime} (x), 
y (x), x]\ ,
\label{S2eq1b}
\end{equation}
where$ H_{nso} [y^{\prime} (x), y (x), x]$ is given by 
Equation (\ref{S1eq12}) and it requires $B (x) = 0$.  The 
auxiliary condition becomes 
$\bar v^{\prime \prime} (x) = x \bar v (x)$, and by
using it, the original Airy equation is obtained from 
$L_{nso} [y^{\prime} (x), y (x), x]$.   

We may also include the following mixed Lagrangian 
\begin{equation}
L_{mo} [y^{\prime} (x), y (x)] = {1 \over 2} q y^{\prime} 
(x) y (x)\ ,
\label{S2eq1c}
\end{equation}
where $q$ is an arbitrary constant.  The resulting gauge
function is 
\begin{equation}
\phi_o (x)  = {1 \over 8} q y^2 (x)\ .
\label{S2eq1d}
\end{equation}

This is the gauge function for the Airy equation, and it is also
the gauge function for the ODEs of the form  $\hat D_o y (x) = 0$
(see Equation (\ref{S1eq10b})).

\subsection{Bessel Equations}

Let us consider a general form of Bessel equations and write 
$\hat D y(x) = 0$ as
\begin{equation}
y^{\prime \prime} (x) + {{\alpha} \over {x}} y^{\prime} (x) + 
\beta \left ( 1 + \gamma {{\mu^2} \over {x^2}} \right ) y (x)
= 0\ ,
\label{S2eq1}
\end{equation}
where $B(x) = \alpha / x$ and $C(x) = \beta (1 + \gamma \mu^2 
/ x^2$).  In addition, $\alpha$, $\beta$, $\gamma$ and $\mu$ 
are constants, and their specific values determine the four 
different types of Bessel equations (see Table \ref{tab1}).  

Based on the above definitions of $B(x)$ and $C(x)$, these 
functions have singularities at $x=0$ (regular) and $x=\infty$ 
(irregular); thus, $x\ \epsilon\ (0, \infty)$; thus, the functions 
are only smooth in this interval, and this is consistent with 
our definition given in Section 1.  In other words, the results 
presented in this section are only valid inside this interval.

The solutions to the Bessel equations are given as the power 
series expansions around the regular singular point $x=0$.  
The obtained solutions to different Bessel equations are 
summarised in Table 2.  All solutions are converging when 
$x \rightarrow \infty$; however, only some are finite in the 
entire range $x\ \epsilon\ (0, \infty)$ but others become 
infinite when $x \rightarrow 0$ [3].

\begin{table}
\caption{Values of the constants $\alpha$, $\beta$, $\gamma$ 
and $\mu$ in Equation (\ref{S2eq1}) corresponding to the four 
types of Bessel equations.}\
\centering
\begin{tabular}{l c c c c c } 
\textbf{Bessel Equations} & \boldmath{$\alpha$} & \boldmath{$\beta$}  
& \boldmath{$\gamma$} & \boldmath{$\mu$}\\

 Regular            &  1 &  1 & $-$1 & real or integer\\
 Modified           &  1 & $-$1 &  1 & real or integer\\
 Spherical          &  2 &  1 & $-$1 & $\mu^2 = l (l+1)$\\
 Modified spherical &  2 & $-$1 &  1 &  $\mu^2 = l (l+1)$\\

\end{tabular}

\end{table}

Using the results of Propositions 1, 2 and 3, we find the 
standard and non-standard Lagrangians for the Bessel
equations are  
\begin{equation}
L_s [y^{\prime} (x), y (x), x] = {1 \over 2} \left [ 
\left ( y'(x) \right )^2 - \beta \left ( 1 + \gamma 
{{\mu^2} \over {x^2}} \right ) y^2 (x) \right ] 
x^{\alpha}\ ,
\label{S2eq2}
\end{equation}
and
\begin{equation}
L_{ns} [y^{\prime} (x), y (x), x] = H_{ns} [y^{\prime} (x), 
y (x), x]\ x^{-2 \alpha}\ ,
\label{S2eq3}
\end{equation}
where $H_{ns} [y^{\prime} (x), y (x), x]$ is given by Equation 
(\ref{S1eq12}), and that the mixed Lagrangian $L_m$
is defined as
\begin{equation}
L_m [y^{\prime} (x), y (x), x] = {{\alpha} \over 2} 
\left [ y'(x) + {{(\alpha - 1)} \over {2 x}} y(x) \right ] 
y(x) x^{\alpha - 1}\ ,
\label{S2eq4}
\end{equation}
with $\alpha$ being either $1$ or $2$.  According to 
Table \ref{tab1}, $\alpha = 1$ corresponds to regular and 
modified Bessel equations for which $L_m [y^{\prime} 
(x), y (x), x] = y(x) y'(x) / 2$; however, for spherical 
and modified spherical Bessel equations $\alpha = 2$ and
$L_m [y^{\prime} (x), y (x), x] = y(x) y'(x) x + y^2(x) 
/ 2$. 

An interesting result is that $L_m [y^{\prime} (x), y (x), 
x] = y(x) y'(x) / 2$ is also the null Lagrangian because 
\begin{equation}
L_m [y'(x), y(x), x] = L_n [y'(x), y(x), x] = {{d \phi} 
\over {d x}}\ ,
\label{S2eq5}
\end{equation}
with 
\begin{equation}
\phi = {1 \over 8} y^2 (x)\ , 
\label{S2eq6}
\end{equation}
being the gauge function for the regular and modified Bessel equations;
it is also the gauge function for the Euler equations because its $b (x) =
1 / x$.  The main reason is that for these equations $B(x) = 1 / x$, 
which gives $B^{\prime} + B^2 (x) = 0$ and reduces significantly 
the derived MLs that are the NLs.  As a result, only in theses cases, can
the gauge functions be defined.   It is interesting that Equation 
(\ref{S2eq6}) can be obtained from Equation (\ref{S1eq10b}) by 
taking $q = 1$.

The auxiliary condition that must supplement $L_{ns} 
[y^{\prime} (x), y (x), x]$ is given by 
\begin{equation}
\bar v^{\prime \prime} (x) + {{\alpha} \over {x}} \bar v^{\prime} 
(x) = - \beta \left ( 1 + \gamma {{\mu^2} \over {x^2}} \right ) 
\bar v (x)\ ,
\label{S2eq7}
\end{equation}
and this condition is required in order to derive the original 
Bessel equations from the E--L equation (see Proposition 3). 
  
The solutions presented in Table \ref{tab2} are the two linearly independent 
solutions for the Bessel equations given as the special functions of
mathematical physics.  The notation used for these solutions is standard 
(e.g., [1,2,3]), which means that $J_{\mu}(x)$ is the Bessel function 
of the first kind, $J_{-\mu}(x)$ is the independent second solution, 
$Y_{\mu}(x)$ is the Bessel function of the second kind or the Neumann 
function, $I_{\mu}(x)$ is the modified Bessel function of the first 
kind, $I_{-\mu}(x)$ is the independent second solution and $K_{\mu}(x)$ 
is the modified Bessel function of the second kind or the modified 
Neumann function.  In addition, $j_{l}(x)$ and $y_{l}(x)$ are the 
spherical Bessel functions, and $i_{l} (x)$ and $k_{l} (x)$ are the 
modified spherical Bessel functions.  According to the results of 
Corollary 6, the derived NSLs  can be expressed in terms of the pairs 
of these solutions, and such explicit dependence of the NSLs  on the
solutions of the original ODEs is a new phenomenon in the calculus 
of variations.

\begin{table}
\caption{The linearly independent solutions of the four Bessel 
equations (see Table 1).  The standard notation commonly adopted 
in textbooks and monographs of mathematical physics is used for 
these solutions.}
\centering
\begin{tabular}{l c c c } 
\textbf{Bessel Equations} & \textbf{Solutions} & \boldmath{$\mu$} 
\textbf{or} \boldmath{$l$}\\

 Regular            &  $J_{\mu}(x)$, $J_{-\mu}(x)$ & real\\
 Regular            &  $J_{\mu}(x)$, $Y_{\mu}(x)$  & integer\\
 Modified           &  $I_{\mu}(x)$, $I_{-\mu}(x)$ & real\\
 Modified           &  $I_{\mu}(x)$, $K_{\mu}(x)$  & integer\\
 Spherical          &  $j_{l}(x)$,   $y_{l}(x)$    & integer\\
 Modified Spherical &  $i_{l}(x)$,   $k_{l}(x)$    & integer\\

\end{tabular}

\end{table}

Among the solutions listed in Table \ref{tab2}, the SFs $I_{\mu} (x)$, 
$I_{-\mu}(x)$, $K_{\mu}(x)$, $i_{l}(x)$ and $k_{l}(x)$ are 
non-oscillatory; however, all other special functions listed above 
are oscillatory.  The superpositions of the solutions $y(x)$ and
$\bar v(x)$ for each Bessel equation do not lead to any discontinuity 
in the above SLs  and NSLs .  However, it must be noted that some of 
the solutions given in Table \ref{tab2} become zero at $x = 0$ but this also 
does not result in discontinuities for the derived NSL because $x 
\in (0, \infty)$. 

\subsection{Legendre Equations}

There are the regular and associated Legendre equations, and 
the latter can be written as
\begin{equation}
y^{\prime \prime} (x) - {{2x} \over {(1-x^2)}} y^{\prime} (x) + 
\left [ {{l (l+1)} \over {(1-x^2)}} - {{m^2} \over {(1-x^2)^2}}
\right ] y (x) = 0\ ,
\label{S2eq8}
\end{equation}
where $l$ and $m$ are constants, and when $m = 0$ the above 
equation becomes the regular Legendre equation (see Table \ref{tab3}). 
Comparing the above Legendre equation to $\hat D y(x) = 0$, 
we obtain $B(x) = - 2 x / (1 -x^2)$ and $C(x) = l (l+1) / 
(1 - x^2) - m^2 / (1 - x^2)^2$, which show that the solutions 
presented below are only valid within the range  $x\ \epsilon\ 
(- 1, + 1)$ as only this interval the functions are smooth.   

For the regular Legendre equation, the power series solutions are 
calculated either at one of the regular singular points $x = \pm 1$ 
[1,2,3].  The two linearly independent solutions are the 
Legendre functions of the first kind or the Legendre polynomials 
$P_l (x)$, which are oscillatory within $x\ \epsilon\ (- 1, + 1)$, 
and the Legendre functions of the second kind $Q_l (x)$ that are 
singular at $x = \pm 1$ (see Table 3); note that $Q_l(x)$ can be 
expressed in terms of $P_l(x)$; nevertheless, the solutions remain 
linearly independent$^{1}$.  The power series solutions calculated at 
one of the regular singular points diverge at $x = \pm 1$ unless $l$ 
is chosen to be an integer, which terminates the series, and finite 
Legendre polynomials of order $l$ are obtained [3]; therefore, 
in physical applications $l$ is a positive integer.  

Because of the above constraints on the solutions, the original 
definitions of $B(x)$ and $C(x)$ given in Section \ref{sec1} are not 
valid, and as a result, the Legendre equations formally do not 
belong to $\mathcal{Q}_{sf}$.  However, since the equations 
are of the form of the ODEs of $\mathcal{Q}_{sf}$, let us make 
adjustments to the definitions of $B(x)$ and $C(x)$, so that our 
procedure of deriving the Lagrangians applies to the Legendre 
equations as well.  It must be noted that the adjustments are 
only valid for the equations considered in this subsection.

We now present the SL only for the associated Legendre equations 
because the SL corresponding to the regular Legendre equation is 
obtained by taking $m = 0$.  However, the NSL and $L_m$ have 
the same forms for both regular and associated Legendre equations.
Following Propositions 1 and 2, we~find the following Lagrangians 
\[
L_s [y^{\prime} (x), y (x), x] = {1 \over 2} \left [ 
y'(x) \right ]^2 (1 - x^2)
\]
\begin{equation}
\hskip0.75in - \left [ {{l (l+1)} \over {(1-x^2)}} - 
{{m^2} \over {(1-x^2)^2}} \right ] y^2 (x) (1 - x^2)\ ,
\label{S2eq9}
\end{equation}
\begin{equation}
L_{ns} [y^{\prime} (x), y (x), x] = H_{ns} [y^{\prime} (x), 
y (x), x]\ (1 - x^2)^{-2}\ ,
\label{S2eq10}
\end{equation}
where $H_{ns} [y^{\prime} (x), y (x), x]$ is given by Equation 
(\ref{S1eq12}), and the mixed Lagrangian is given by 
\begin{equation}
L_m [y^{\prime} (x), y (x), x] = - \left [ x y'(x)
+ {1 \over 2} y (x) \right ] y(x)\ ,
\label{S2eq11}
\end{equation}
which shows that the gauge function cannot be defined for
this mixed Lagrangian.  It is also interesting to note that the 
mixed Lagrangians for the spherical Bessel and Legendre 
equations have $L_{m} [y^{\prime} (x), y (x), x]$ that 
differs only by the sign. 

Proposition 3 shows that substitution of $L_{ns} [y^{\prime} (x), 
y (x), x]$ into the E--L equation does not result in the original
Legendre equations unless the following auxiliary condition 
\begin{equation}
\bar v^{\prime \prime}(x) - \left [ {{2x} \over {(1-x^2)}} \right ]\ 
\bar v^{\prime}(x) = - \left [ {{l (l+1)} \over {(1-x^2)}} - {{m^2} 
\over {(1-x^2)^2}} \right ]\ \bar v(x)\ , 
\label{S2eq12}
\end{equation}
is taken into account.\\ 

\begin{table}
\caption{The linearly independent solutions of the regular and 
associated Legendre equations.  The~standard notation commonly 
adopted in textbooks and monographs of mathematical physics is
used for these solutions.} 
\centering
\begin{tabular}{l c c c } 
\textbf{Legendre Equations} & \textbf{Solutions} & \emph{\textbf{m}}\\

 Regular            &  $P_{l}(x)$,     $Q_{l}(x)$     & 0\\
 Associated         &  $P_{l}^{m}(x)$, $Q_{l}^{m}(x)$ & $-$l,..,0,..,l\\

\end{tabular}

\end{table}

There are two linearly independent solutions of the associated 
Legendre equation, the associated Legendre functions of the first 
kind $P_l^m (x)$, which are related to $P_l (x)$, and the associated 
Legendre functions of the second kind $Q_l^m (x)$, which are related 
to $Q_l (x)$; in the case in which $l$ is an integer, $P_l^m (x)$ is called 
the associated Legendre polynomials.  The well-known property of the 
Legendre functions is the fact that all $P_l^m (x)$ with $m > 0$ can 
be generated from $P_l (x)$, which can be built recursively from 
$P_0 (x) = 1$ and $P_1 (x) = x$; the same is true for $Q_l^m (x)$.  
It must be pointed out that $P_l^m (x)$ are oscillatory within $x\ 
\epsilon\ (- 1, + 1)$ and $Q_l^m (x)$ are non-oscillatory.  These 
properties of the solutions are important for expressing the SL, NSL 
and $L_m$ in terms of the superpositions of linearly independent 
solutions $y_1(x)$ and $y_2(x)$, as shown by Corollary 6.

\subsection{Hermite Equation}

The Hermite equation can be written as 
\begin{equation}
y^{\prime \prime} (x) - x y^{\prime} (x) + n y (x) = 0\ ,
\label{S2eq13}
\end{equation}
where $n$ is any integer.  Comparing this equation to 
$\hat D y(x) = 0$, we find $B(x) = - x$ and $C(x) = n$.
The~range of validity of the Hermite equation is $x\ 
\epsilon\ (0, \infty)$.

The Hermite equation has only one singular point at 
infinity and this point is the irregular singular point.  
Despite being irregular, the power series solutions can 
still be obtained about this point because of its location 
at the end of the interval for $x$.  Another possibility 
is to construct the power series solutions about any other 
point, say $x = 0$, which would be an ordinary point.  The 
two power series solutions obtained around the irregular 
singular point located at $x = \infty$ give two linearly
independent solutions: the Hermite functions of the 
first kind or the Hermite polynomials $H_n (x)$ if, 
and only if, $n \geq 0$ is an integer, and the Hermite 
functions of the second type $h_n (x)$ .  

The explicit forms of the standard, non-standard and 
mixed Lagrangian for the Hermite equation~are
\begin{equation}
L_s [y^{\prime} (x), y (x), x] = {1 \over 2} \left [ 
\left ( y'(x) \right )^2 - n y^2 (x) \right ] e^{-x^2/2}\ ,
\label{S2eq14}
\end{equation}
\begin{equation}
L_{ns} [y^{\prime} (x), y (x), x] = H_{ns} [y^{\prime} (x), 
y (x), x] e^{x^2}\ ,
\label{S2eq15}
\end{equation}
where $H_{ns} [y^{\prime} (x), y (x), x]$ is given by Equation 
(\ref{S1eq12}), and mixed Lagrangian is defined as 
\begin{equation}
L_m [y^{\prime} (x), y (x), x] = - {1 \over 2} \left [ 
x y'(x) + {1 \over 2} ( 1 - x^2) y (x) \right ] 
y(x) e^{-x^2/2}\ ,
\label{S2eq16}
\end{equation}
and the gauge function cannot be defined, which means
that $L_m [y^{\prime} (x), y (x), x]$ is not the null
Lagrangian and that the gauge function cannot be 
determined.  

It is also seen that the terms $x y(x) y'(x) + y^2(x)/2$ that 
appeared in the mixed Lagrangians for the spherical Bessel 
and Legendre equations are also present in the mixed  
Lagrangians for the Hermite equation.  To obtain the original 
Hermite equation, the Lagrangian must be substituted into 
the E--L equation, and according to Proposition 3 the following 
auxiliary condition 
\begin{equation}
\bar v^{\prime \prime}(x) + x \bar v^{\prime}(x) = - n \bar 
v(x)\ ,
\label{S2eq17}
\end{equation}
must be used.

\subsection{Discussion of Applications}

Having obtained the standard and non-standard Lagrangians for the 
Airy, Bessel, Legendre and Hermite equations is equivalent to showing 
that the equations can be derived from the Lagrangian formalism based 
on these Lagrangians [24,34].  Similarly, we may find the SLs  and NSLs 
for all other ODEs of $\mathcal{Q}_{sf}$, and establish the Lagrangian 
formalism for all considered ODEs.  The obtained results are important 
in theoretical physics, since their main equations are typically derived 
from given Lagrangians.  Our results may also be useful in applied 
mathematics and engineering where the ODEs $\mathcal{Q}_{sf}$ 
are commonly used.  

There are advantages of having the Lagrangian formalism for the ODEs 
of $\mathcal{Q}_{sf}$, and they include using the derived Lagrangians
to study the group structure underlying the SLs  and NSLs, and~their 
symmetries (possibly some new ones), as well as finding relationships 
between the Lagrange formalism and the Lie group approach (e.g., [1,6]), 
which introduces the special functions by using irreducible representations 
of some Lie groups; these topics are out the scope of this paper but they 
will be investigated in the future.

We also derived the mixed Lagrangians for $\hat D_o y (x) = 0$ and for 
$\hat Dy (x) = 0$.  For the former, we demonstrated that the mixed 
Lagrangians depend on an arbitrary constant and that the gauge functions 
can be defined for all these MLs since they are equivalent to the NLs. 
For the latter, we showed that the MLs depend on the function $B(x)$ 
but they are independent of $C(x)$.  Because of this dependence on 
$B(x)$, the MLs become zero when $B(x) = 0$.  Other special cases 
of $L_{m} [y'(x), y(x), x] = 0$ are also possible, and we now list the 
required conditions: (i) $y(x) = 0$ and $B(x) = 0$; (ii) $y(x) = 0$ but 
$B(x) \neq 0$; (iii) $y(x) \neq 0$ but $B(x) = 0$; and $G_{m} [y'(x), 
y(x), x] = 0$, which is only satisfied when $y(x)$ and $B(x)$ are related 
by $y(x) = C_0 \exp [- (\int^x B (\tilde x)$ $d \tilde x) / 2] / \sqrt{B(x)}$, 
and with $C_0$ being the integration constant.  

The existence of the MLs for the ODEs of $\mathcal{Q}_{sf}$ is an 
interesting result, which shows that there is a family of Lagrangians 
that gives null contributions to calculus of variations by fully solving 
the E--L equation.  Specific applications of the MLs and NLs are to study 
symmetries of physical systems described by the Lagrangians, and 
their other effects will be the subject of future explorations.  An important 
result is that the only MLs obtained for the Airy equation, and regular 
and modified Bessel equations become the NLs, and that only for 
these two equations can the gauge functions be properly defined.

\section{Conclusions}

We considered linear second-order ODEs with non-constant coefficients 
that are commonly used in applied mathematics, physics and engineering.  
We selected the ODEs with non-constant coefficients whose solutions are 
the special functions of mathematical physics, and denoted them as 
$\mathcal{Q}_{sf}$.  We~established the Lagrange formalism for these 
equations.   Since the original ODEs were known, our main objective was 
to derive the Lagrangians corresponding to these equations.  This required 
solving the inverse calculus of the variations problem, and we developed novel 
methods for solving it.  These methods allowed for the derivation of standard 
and non-standard Lagrangians, and also the mixed Lagrangians that make 
null contributions to the Lagrangian formalism but become full solutions 
of the Euler--Lagrange equation.  We identified the equations for which the 
mixed Lagrangians are the null Lagrangians with the properly defined gauge 
functions.  An interesting result is that for most ODEs of $\mathcal{Q}_{sf}$,
the gauge functions cannot be defined.  We also showed that the derived 
non-standard Lagrangians form a new family of Lagrangians.  

The dependence of non-standard Lagrangians on an additional dependent
variable requires that the E--L equation is amended by the auxiliary 
condition, which is again, a new phenomenon in calculus of variations.  
To present the effects of these new phenomena, we considered the 
Helmholtz conditions and investigated their validity. The obtained 
results clearly showed that there are ODEs in the set $\mathcal{Q}_{sf}$ 
for which the non-standard Lagrangians can be found, despite the fact 
that the Helmholtz conditions are violated.  This result may have 
profound implications on the development of the Lagrangian formalism 
based on non-standard Lagrangians.   

We considered specific applications of our results to the Airy, Bessel, 
Legendre and Hermite equations.  In these applications, we constructed 
the standard, non-standard and mixed Lagrangians for each one of these 
equations, and discussed the similarities and differences between the 
resulting Lagrangians.  The presented results demonstrate that the 
Lagrangian formalism is well-established for the ODEs of $\mathcal{Q}_{sf}$,
which means that there is a powerful and robust method of obtaining 
the ODEs whose solutions are given in terms of the special functions of 
mathematical physics.  Moreover, our~results show that the derived 
mixed Lagrangians become the null Lagrangians only for the Airy 
(and other similar) equations, and also for the regular and modified 
Bessel equations.

Finally, let us briefly summarise important open problems resulting from 
this paper.  The presented methods of finding standard and non-standard 
can be extended to other ODEs both homogeneous and inhomogeneous.  
The obtained mixed (null) Lagrangians can be used to investigate 
symmetries of physical systems described by such Lagrangians. 
Since only one family of non-standard Lagrangians was explored, other 
families known from the literature [17,20,25] should also be investigated 
to determine whether all these Lagrangians require the auxiliary conditions. 
The fact that the Euler--Lagrange equation does not give directly the original 
equation is a new phenomenon that needs additional studies. 
Some non-standard Lagrangians violate the Helmholtz conditions; therefore,
a generalisation of these conditions, so they apply to all Lagrangians, is required. 
These open problems will be explored in separate papers.  
%

{\bf Acknowledgments:} We are indebted to anonymous referees for carefully reading our 
paper and providing many detailed comments and suggestions that helped us to significantly 
improve the revised version of this paper.  This~work was partially supported by the Alexander 
von Humboldt Foundation (Z.E.M.) and by the LSAMP Program at the University of Texas 
at Arlington (M.R.-F.). 



\end{document}